# Q-tensor model of twist-bend and splay nematic phases

Martin Čopič[1,2], and Alenka Mertelj[1]

[1]*J. Stefan Institute, SI-1000 Ljubljana, Slovenia*

[2]*Faculty of Mathematics and Physics, University of Ljubljana, SI-1000 Ljubljana, Slovenia*

(Date: 2019-10-03)

The twist-bend nematic phase ($N_{TB}$) is characterized by a conically twisting director and by a dramatic softening of the bend elastic constant before the transition to $N_{TB}$ phase. In the recently found splay nematic phase ($N_S$) the splay elastic constant tends to zero, resulting in a splay modulation perpendicular to the director. We model both phases with a single Q-tensor free energy including a term that breaks the degeneracy between the splay and bend elastic constant, and a flexoelectric term coupling the divergence of the Q-tensor with polarization. $N_{TB}$ or $N_S$ phase is obtained by a change of sign of one elastic parameter. Measured elastic constants show that $N$-$N_{TB}$ transition is mainly driven by the increase of the nematic order, while the $N_S$ transition is due to flexoelectric coupling.

**Subject Areas:** Soft Matter, Materials Science, Phase transitions

## I. INTRODUCTION

In recent years it has been found that nematic liquid crystals exhibit transitions to new phases that are still nematic but are periodically modulated. The first such phase is the twist-bend nematic ($N_{TB}$) in which the nematic director helically twists about an axis parallel to the director in the normal nematic (*N*) phase [1–4]. The helix can be either right or left handed, so the transition breaks the chiral symmetry of the normal *N* phase. The molecules that form this phase are flexible dimers of some common nematogens that tend to have bent relaxed shape. A characteristic property of these materials is that the bend elastic constant in the *N* phase is unusually small and strongly decreases upon approaching the N$_{TB}$ phase. The softening of the bend elastic constant was also obtained in the numerical molecular modeling of Cestari et al [5].

Recently we found another nematic-nematic transition [6] where in the normal high temperature N phase the splay elastic constant becomes very small, so that the low temperature phase must exhibit spontaneous splay. As it is impossible to fill space with homogeneous splay, this N$_S$ phase must be periodically splayed in the direction perpendicular to the nematic director.

There are several models that have been proposed for the $N_{TB}$ phase. This phase was theoretically predicted by I. Dozov [7], who observed that if for some reason the bend elastic constant goes to zero, a modulated phase must result. In his model two possible states are predicted, either a helically twisted phase where bend is accompanied by twist, or a splay-bend phase ($N_{SB}$) in which the bend is planar and accompanied by some splay. The modulation vector is parallel to the average director. This $N_{SB}$ phase should occur for sufficiently large twist elastic constant. In Dozov's Landau type model the deformed state is stabilized by terms with second derivatives of the director.

Based on an observation by R. Meyer [8] that the coupling of the flexoelectric polarization, that is electric polarization that arises as a result of bend or splay deformation, could make the nematic state unstable, Shamid et al. [9] proposed a Landau-de Gennes model with coupled bend deformation of the director and polarization and showed that this can describe the $N_{TB}$ phase.

Recently, L. Longa [10] analyzed a free energy with the primary nematic order parameter a traceless second rank tensor $Q_{ij}$. For the normal $N$ phase $\mathbf{Q} = S\left(\mathbf{n} \otimes \mathbf{n} - \frac{1}{3}\mathbf{I}\right)$, where $S$ is the degree of nematic order. In the free energy Longa included two isotropic quadratic terms in the gradient of the order tensor $\nabla \mathbf{Q}$ and bilinear coupling of $\nabla \mathbf{Q}$ with a polarization vector $\mathbf{P}$ plus a number of other terms. Longa stresses that $\mathbf{P}$ does not have to be the electric polarization, in which case the coupling term describes flexoelectric effect, but can also be a vector associated with the shape of the molecules. In the case of bent molecules this is a vector perpendicular to the molecular long axis in the plane of the bend. In the analysis he finds several modulated phases including the $N_{TB}$, but not the $N_S$.

In this work we present a Landau-de Gennes type model also based on Q-tensor expansion and coupling to polarization vector. In the free energy used by Longa the bend and splay elastic constants are degenerate. In our model we add terms that break this degeneracy and show that by just changing the sign of the degeneracy breaking term both the $N_{TB}$ and $N_S$ phases can be obtained. We also present experimental data on elastic constants that shed some additional light on the source of the instability leading to the transition.

## II. THEORY

### A. Nematic elastic constants

The free energy density describing isotropic-nematic phase transition in terms of Q-tensor was given by de Gennes [11]

$$f = \frac{A}{2}Q^2 + \frac{B}{3}Q^3 + \frac{C}{4}(Q^2)^2 + \frac{1}{2}L_1 Q_{ij,k} Q_{ij,k} + \frac{1}{2}L_2 Q_{ij,j} Q_{ik,k} \quad (1)$$

The constant A is assumed to become negative at a certain temperature $T_c$, and for the usual N phase with $S > 0$ the constant $B$ is negative, so that the transition is first order and $\mathbf{Q}$ uniaxial. The two elastic terms in $\nabla \mathbf{Q}$ are the only isotropic volume ones, second order in $\mathbf{Q}$. When the nematic elastic constants in the free energy in terms of the director are expressed in deGennes constants, the splay constant $K_1$ and the bend constant $K_3$ are equal, while the twist constant $K_2$ is different. To break this degeneracy, it is necessary to add to free energy (1) new terms:

$$f = \frac{A}{2}Q^2 + \frac{B}{3}Q^3 + \frac{C}{4}(Q^2)^2 + \frac{1}{2}L_1 Q_{ij,k} Q_{ij,k}$$
$$+ \frac{1}{2}L_2 Q_{ij,j} Q_{ik,k} + L_{31} Q_{ij} Q_{ik,j} Q_{kl,l} + L_{32} Q_{ij} Q_{ik,k} Q_{jl,l} \quad (2)$$
$$+ L_{33} Q_{ij} Q_{ik,l} Q_{jl,k} + L_{34} Q_{ij} Q_{ik,l} Q_{jk,l} + L_{35} Q_{ij} Q_{ik,l} Q_{kl,j}$$

There are two other similar terms that are linearly dependent or differ by a divergence (surface term) from the ones given above.

The Frank-Oseen usual nematic constants are related to the above in the following way:

$$K_1 = S^2 (2L_1 + L_2) + \frac{2}{3}S^3 (-L_{31} + 2L_{32} + L_{33} + 2L_{34} - L_{35})$$

$$K_2 = 2S^2 L_1 + \frac{2}{3}S^3 L_{33}$$

$$K_3 = S^2 (2L_1 + L_2) + \frac{2}{3}S^3 (2L_{31} - L_{32} + L_{33} - L_{34} + 2L_{35})$$
(3)

Four of the five terms $L_{3i}$ break the degeneracy of between $K_1$ and $K_3$. It is important to note that their contribution is proportional to $S^3$.

### B. Model of $N_{TB}$ and $N_S$ phase

We construct the free energy density that describes the phase transition from the normal homogeneous N phase to the spatially modulated $N_{TB}$ or $N_S$ phase following the idea, developed in Ref. [9], that the elastic instability results from the flexoelectric coupling between the gradient of the director and polarization $\mathbf{P}$. In our case the only such term is with the divergence of $\mathbf{Q}$:

$$f = \frac{1}{2}L_1 Q_{ij,k} Q_{ij,k} + \frac{1}{2}L_2 Q_{ij,j} + L_{31} Q_{ij} Q_{ik,j} Q_{kl,l} - \gamma P_i Q_{ij,j}$$
$$+ \frac{1}{2}t P_i P_i + \frac{1}{2}d P_{i,j} P_{i,j}$$
(4)

We only include one term that breaks the splay-bend degeneracy. Within the approximation that $\mathbf{Q}$ remains uniaxial in the deformed phase, all other terms lead to the same results. The parameter $t$ in the leading term of energy associated with $\mathbf{P}$ contains a positive electrostatic contribution, a positive entropic part increasing with $T$, and a negative molecular packing contribution, so we assume it is an increasing function of $T$. It does not have to become negative at some $T_0$ as assumed in Ref. [9]. The parameter $\gamma$ describes the flexoelectric coupling, and the term $d\, P_{i,j}P_{i,j}$ is the lowest order one that stabilizes the deformed phases. Free energy density (4) is the minimal one to describe the $N$ to $N_{TB}, N_S$ transitions.

### 1. $N_{TB}$ transition

In the case of $N_{TB}$ transition the bend elastic constant is smaller than splay, so according to eq. (3) we take $L_{31} < 0$. We also assume **Q** to remain uniaxial in the $N_{TB}$ phase. This is an approximation as the local symmetry of the $N_{TB}$ or $N_S$ phase is biaxial, but the amount of biaxiality is expected to be small. Namely, if the amount of biaxiality of **Q** is $p$, then the associated free energy density from expression (1) is $|B|S\, p^2$. The ratio $\sqrt{L_1/|B|} \approx 1$ nm, so unless the gradient of **Q** is of the order of 1 nm$^{-1}$, as may occur in the core of a defect, the biaxiality induced by elastic deformations is small. So we take **Q** of the form (Fig. 1a)

$$\mathbf{Q}(z) = S\left(\mathbf{n}\otimes\mathbf{n} - \frac{1}{3}\mathbf{I}\right) \quad (5)$$
$$\mathbf{n}(z) = (\sin\theta\cos kz, \sin\theta\sin kz, \cos\theta).$$

The magnitude of the nematic order parameter $S$ is governed by the first three terms in Eq. (1). In the following we assume that it is a given function increasing with decreasing $T$ and is unaffected by the transition to the modulated phase.

Polarization must also depend only on $z$ coordinate:

$$\mathbf{P}(z) = [p_1(z), p_2(z), p_3(z)]. \quad (6)$$

Putting (5) and (6) into (4), the obtained expression for free energy is a functional of $\mathbf{P}(z)$. The corresponding Euler-Lagrange equations have the solution

$$\mathbf{P} = \frac{\gamma k}{2(dk^2+t)}\sin(2\theta)(-\sin kz, \cos kz, 0). \quad (7)$$

Inserting **P** into $f$ and expanding $f$ to $\theta^4$ (or introducing $\sin\theta$ as a new variable), we get a reduced free energy density

$$f_Q = \frac{1}{2(t+dk^2)}\left\{c_b\left(t - \frac{\gamma^2}{c_b} + dk^2\right)\theta^2 \right. \\ \left. + \left[(2L_1S^2 - c_b)(t+dk^2) + \gamma^2\right]\theta^4\right\}, \quad (8)$$

where $c_b = 2L_1S^2 + L_2S^2 - 4/3|L_{31}|S^3$. For $t < t_c$, where $t_c = \gamma^2/c_b$, the N phase with $\theta = 0$ is stable. The effective bend elastic constant

$$K_{3eff} = c_b - \frac{\gamma^2}{t} = (2L_1+L_2)S^2 - \frac{4}{3}|L_{31}|S^3 - \frac{\gamma^2}{c_b} \quad (9)$$

goes to 0 at $t_c$. Note that the instability can also be driven by the negative value of $L_{31}$ and increasing $S$ instead of decreasing $t$.

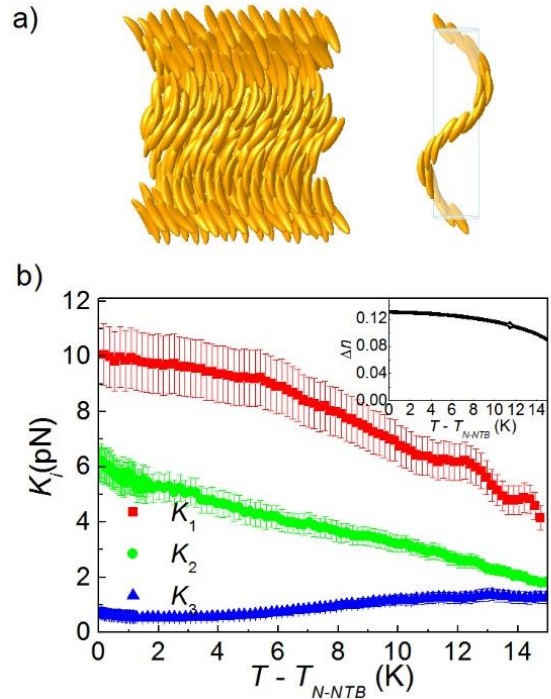

Figure 1: (color online) (a) Scheme of the twist-bend nematic phase. (b) Temperature dependence of the splay (red squares), the twist (green circles), and the bend (blue triangles) elastic constants in the nematic phase above the $N$-$N_{TB}$ phase transition. Inset: Temperature dependence of the anisotropy of the index of refraction.

For $t<t_c$ the $N_{TB}$ is stable. To find the equilibrium values of $\theta$ and $k$ we have to solve the equations $\frac{df}{d\theta}=0$ and $\frac{df}{dk}=0$. The solution is rather complicated but it can be expanded to the lowest order in $\Delta t = t_c - t$ to get

$$\theta_{TB} = \frac{c_b}{\gamma}\sqrt{\frac{\Delta t}{6c_1}} \text{ and } k = \sqrt{\frac{\Delta t}{3d}} \ . \quad (10)$$

Also to the same order

$$\mathbf{P} = \frac{\gamma k}{2t_c}\sin(2\theta)(-\sin kz, \cos kz, 0). \quad (11)$$

These expressions are very similar to the ones derived by Shamid et al. [9]. Note that the possible choice of sign of $\theta$ relative to $k$ reflects the chiral symmetry breaking in the $N_{TB}$ phase.

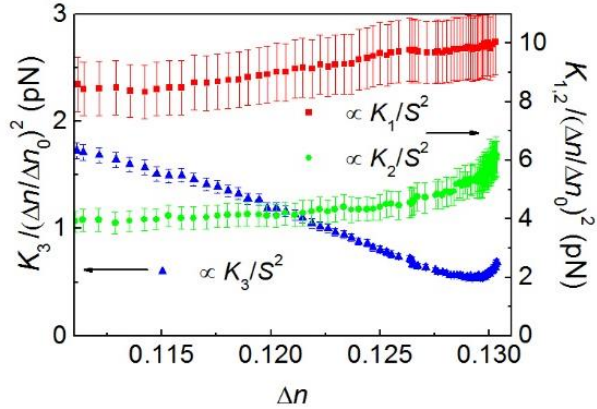

Figure 2: (color online) Dependence of the bend (blue triangles), the twist (green circles) and the splay (red squares) elastic constants divided by $(\Delta n/\Delta n_0)^2$ on $\Delta n$. $\Delta n_0$ is the value of $\Delta n$ at $T_{N\text{-}NTB}$.

We can also look for the conditions under which the $N_{SB}$ phase theoretically predicted by I. Dozov [7] would be stable. In this case the director must be of the form

$$\mathbf{n} = (\sin(\theta\sin kz), 0, \cos(\theta\sin kz)). \quad (12)$$

Putting the corresponding $\mathbf{Q}$ into $f$, integrating over period $2\pi/k$, and expanding to $\vartheta^4$ we again obtain a functional of $\mathbf{P}(z)$ that gives as a solution to the Euler-Lagrange equations

$$\mathbf{P} = \frac{\gamma k\theta}{(t+dk^2)}(\cos kz, 0, \theta\sin 2kz). \quad (13)$$

We put this into $F$ and keep terms up to $\vartheta^4$. For $t>t_c$ the equilibrium conditions give the same softening of the $K_{3eff}$ as before, given in eq. (9). For $t<t_c$ the solution for the equilibrium to the lowest order in $\Delta t$ is

$$\theta_{SB} = \frac{c_b}{\gamma}\sqrt{\frac{2\Delta t}{3(2c_b+|L_{31}|S^3)}} \ , \ k = \sqrt{\frac{\Delta t}{3d}} \ . \quad (14)$$

To find out which structure is the stable one we must compare the equilibrium values of $f$ in both cases. For $N_{TB}$ we get

$$f_{TB} = -\frac{c_b^4 \Delta t^3}{108 L_1 S^2 \gamma^4 d} \ , \quad (15)$$

and for $N_{SB}$

$$f_{SB} = -\frac{c_b^4 \Delta t^3}{54(2c_b+|L_{31}|S^3)d\gamma^4} \ . \quad (16)$$

The $N_{TB}$ phase is stable if

$$2L_1 = K_2 < 2c_b + |L_{31}|S^3 = K_1 + c_b - |L_{31}|S^3. \quad (17)$$

This implies a rather large twist constant so we do not expect the $N_{SB}$ phase to be observed.

### 2. $N_S$ transition

Now we look at the transition to the splay nematic phase $N_S$, where the deformation modulation is perpendicular to $\mathbf{n}$ in the N phase. As in this case the splay constant is soft and leads to instability, we take $L_{31}>0$. We again assume that $\mathbf{Q}$ is uniaxial, so $\mathbf{n}$ and $\mathbf{P}$ must have the form (Fig. 3a)

$$\mathbf{n}(x) = (\sin(\theta\sin kx), 0, \cos(\theta\sin kx))$$
$$\mathbf{P}(x) = (p_1(x), 0, p_3(x)) \quad (18)$$

As before, putting $\mathbf{Q}$ and $\mathbf{P}$ into $f$, integrating over period to get $F = \int f\, dx$, expanding $F$ to $\vartheta^4$, and solving $\frac{\delta F}{\delta P_i} = 0$, we get

$$\mathbf{P}(x) = \left( \frac{\gamma\theta^2 k \sin 2kx}{t + 4k^2 d}, 0, \frac{\gamma\theta k \cos kx}{t + k^2 d} \right) \quad (19)$$

We insert $\mathbf{P}(x)$ back to $F$ and solve $\frac{\partial F}{\partial \theta} = 0$ and $\frac{\partial F}{\partial k} = 0$.

For $t > t_c = \frac{\gamma^2}{c_s}$, where $c_s = 2L_1 S^2 + L_2 S^2 - \frac{4}{3} L_{31} S^3$, we have the normal $N$ phase with $\vartheta = 0$. The effective splay constant is

$$K_{1eff} = c_s - \frac{\gamma^2}{t} \quad (20)$$

It goes to 0 at $t = t_c$.

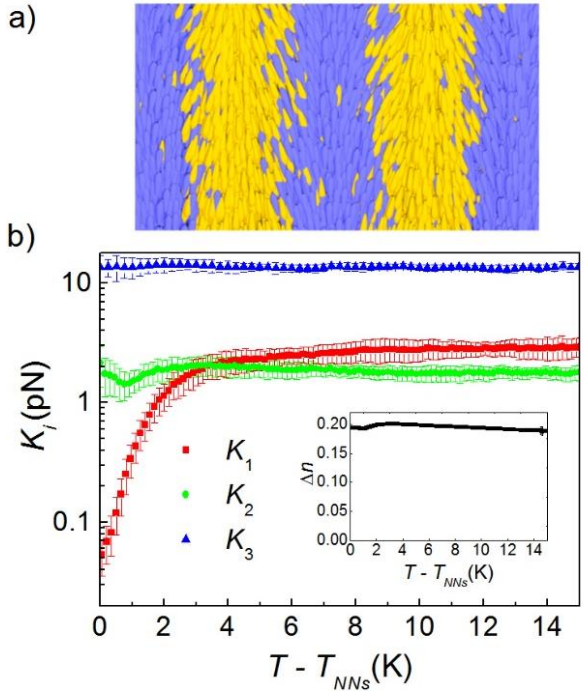

Figure 3: (color online) (a) Scheme of the splay nematic phase. (b) Temperature dependence of the splay (red squares), the twist (green circles), and the bend (blue triangles) elastic constants in the nematic phase above the $N$-$N_s$ phase transition [6]. Inset: Temperature dependence of the anisotropy of the index of refraction.

For $t < t_c$, the modulated phase $N_S$ is stable. To the lowest order in $\Delta t = t_c - t$ we get

$$\theta_s = \frac{c_s}{\gamma} \sqrt{\frac{2\Delta t}{3d(2c_s + b)}}, \quad k = \sqrt{\frac{\Delta t}{3d}},$$

$$\mathbf{P} = \frac{\sqrt{2}}{3} \frac{c_s^2 \Delta t}{\gamma^2 \sqrt{d(2c_s + L_{31} S^3)}} (0, 0, \cos kx) \quad (21)$$

We also checked that all the obtained solutions are minima of $F$ by showing that the Hessian matrix is positive definite.

### III. EXPERIMENT

By dynamic light scattering (DLS) we measured the elastic constants of CB9CB exhibiting $N_{TB}$ transition (Fig. 1) and of RM734 with $N_S$ transition [6]. In DLS we observe the thermal fluctuations of the director. From these measurements, we obtain the temperature dependence of the elastic constants and some viscosity coefficients. We measure the scattered intensity and relaxation rates of eigenmodes of orientational fluctuations. By choosing the correct scattering geometry, pure modes can be measured with the intensities $I_i \propto (\Delta\varepsilon_{opt})^2 / K_i q^2$, and the relaxation rates $1/\tau_i = K_i q^2 / \eta_i$, where $i = 1,2,3$ denote splay, twist and bend and $q$ is the scattering vector [11]. While twist viscosity equals rotational viscosity $\gamma_1$, $\eta_2 = \gamma_1$, the values of bend and splay viscosities are affected by backflow, $\eta_1 = \gamma_1 - \alpha_3^2 / \eta_b$ and $\eta_3 = \gamma_1 - \alpha_2^2 / \eta_c$, where $\alpha_i$ are Leslie viscosity coefficients and $\eta_{b,c}$ Miesowicz viscosities [11]. In the case of the splay, the reduction of orientational viscosity due to the backflow is usually small (of the order of a few %), so $\eta_1 \approx \gamma_1$. The temperature dependence of the anisotropy of dielectric tensor at optical frequencies $\Delta\varepsilon_{opt}$ was obtained from measurements of the optical anisotropy $\Delta n$. With this method, the temperature dependence of elastic constants is obtained, but not their absolute values. The fact that $\Delta n$ is proportional to $S$ is also important in the following analysis.

Let us first look at the results for the CB9CB. Fig. 1b shows the temperature dependence of the elastic constants, with the inset showing $\Delta n(T)$. As our method does not give absolute values of the elastic constants, the absolute values were obtained using the values for $K_1$ and $K_3$ at $T - T_{NI} = -6.8$ K measured by dielectric

Frederiks transition [12,13]. The absolute values of $K_2$ was obtained from the ratio between twist and splay relaxation rates assuming $\eta_1 \approx \gamma_1$. The bend constant decreases in a broad temperature interval and becomes small close to the transition. The increase just before the transition has been observed also in other compounds [12–15], but is currently not well understood. Fig. 2 shows $K_i/\Delta n^2$ as a function of $\Delta n$. As $\Delta n$ is proportional to $S$, according to eq.(3), this plots should be linear functions with the slope given by the coefficients $L_{3i}$. For splay and twist they are increasing, while for bend it is strongly decreasing, signifying a large negative contribution the $L_{3i}$ coefficients. This shows that in the case of CB9CB the main part of the softening of $K_3$ is due to the increase of $S$.

Now let us look at the $N_S$ case. Fig. 3 shows the elastic constants and $\Delta n$ vs. $T$ for RM734. Within 15 degrees of the $N_S$ transition $\Delta n$ is nearly saturated, so we can't do the same plot as in Fig. 2. Now it is the splay constant that becomes very small at the $N_S$ transition. Note also that the $T$ dependence of $K_1$ is qualitatively different from the dependence of $K_3$ in the $N_{TB}$ case. $K_1$ is nearly constant to within a few degrees of the transition and then nonlinearly to a small value, consistent with the $\gamma^2/t$ term in eq. (20) driving the transition.

## IV. DISCUSSION

The results of our model in the case of $N_{TB}$ transition are qualitatively the same as in Ref. [9], where the bend and twist vectors, together with flexoelectric coupling to polarization, are used as variables. A similar model using splay as the variable and flexoelectric polarization can describe the $N_S$ transition. In such models it is the temperature dependence of the $\mathbf{P}^2$ coefficient ($t$ in our model) that drives the instability. It is also possible to simply require that the bend (in case of $N_{TB}$) or splay (in case of $N_S$) elastic constant goes to zero, as was done in Dozov's original paper [7].

Our model, using the Q-tensor, describes both types of elastic instabilities, bend and splay, by a Landau-deGennes type free energy with just a change of sign of a single coefficient. The relevant term of the form $\mathbf{Q}(\nabla \mathbf{Q})^2$ breaks the degeneracy of the bend and splay elastic constants that is present in the isotropic part of the free energy. This contribution to $K_1$ and $K_3$ is proportional to $S^3$. Depending on the sign of this term, it causes either bend or splay constant to decrease with increasing $S$ and decreasing $T$. Flexoelectric coupling also renormalizes the splay and bend constants, as seen in Eqs. (9) and (13), and provides and alternative mechanism for elastic constant softening.

Our measurements of the $T$ dependence of the elastic constants show that in the case of $N_{TB}$ transition the main contribution to the softening of $K_3$ is due to the $S^3$ term. This is shown by the approximately linear decrease of $K_3$ over relatively broad $T$ interval. In the case of $N_S$ transition, $S$ is approximately constant in the vicinity of the transition [16], so it can't be the cause of $K_1$ softening. Also, the form of the $T$ dependence of $K_1$ in Eq. (20) shows that decrease of the coefficient $t$ causes $K_1$ softening.

It is worth noting that our model free energy contains only one flexoelectric coefficient $\gamma$. This is the only possible isotropic term and in this model the bend and splay flexoelectric coefficients are equal. Similarly as in the case of elastic constants, this degeneracy is removed by a term of the form $\mathbf{Q}(\nabla \mathbf{Q})\mathbf{P}$ which is allowed in the uniaxial symmetry of the nematic phase.

## V. CONCLUSIONS

We have presented a unified phenomenological Q-tensor model of the phase transition from the usual $N$ phase to the novel $N_{TB}$ and $N_S$ phases. The choice of the phase depends on the sign of the elastic term cubic in $S$. Comparison with the experimental data on elastic constants show that the $N_{TB}$ transition is mainly driven by the negative $S^3$ contribution to $K_3$, while the $N_S$ phase is due to the flexoelectric coupling.

## VI. ACKNOWLEDGMENTS

Authors thank N. Sebastián for providing the data for elastic constants, which were used to obtain absolute values of $K_i$. This work was funded by the Slovenian Research Agency research core funding No. P1-0192 and project No. J7-8267. MČ also thanks the Isaac Newton Institute for Mathematical Sciences for support and hospitality during the programme *The mathematical design of new materials* when work on this paper was undertaken. This work was supported by EPSRC grant number EP/R014604/1.